\documentclass[12pt]{iopart}

\usepackage[colorlinks=true, pdfstartview=FitV, linkcolor=red, citecolor=blue, urlcolor=blue]{hyperref}
\usepackage{bm}
\usepackage{cite}
\usepackage{graphicx}
\begin{document}

\title[Hydrodynamic model of heavy-ion collisions with low momentum components ]{Hydrodynamic model of heavy-ion collisions with low momentum components }

\author{Akihiko Monnai}

\address{Department of General Education, Faculty of Engineering, 
Osaka Institute of Technology, Osaka 535-8585, Japan}
\ead{akihiko.monnai@oit.ac.jp}
\vspace{10pt}
\begin{indented}
\item[]April 2023
\end{indented}

\begin{abstract}
Relativistic heavy-ion collisions suggest that low momentum regions of the observed particle spectra are thermal and hydrodynamic, while medium-high momentum regions are non-thermal and perturbative. In this study, I construct a hydrodynamic model of heavy-ion collisions by cutting off the medium-high momentum contributions and investigate the phenomenological consequences. Numerical simulations indicate that the temperature of the quark matter can be higher at earlier times owing to the modification of the equation of state. It is also suggested that direct photon elliptic flow can be sensitive to the momentum dependence of thermalization.
\end{abstract}

%
%
%
%
%

\section{Introduction}
\label{sec:1}

Heavy-ion collisions at relativistic energies can produce quark-gluon plasma (QGP) \cite{Letessier:2002gp,Rafelski:2003zz,Yagi:2005yb,Wang:2016opj}, a deconfined quantum chromodynamic (QCD) matter that is considered to have filled the early universe. BNL Relativistic Heavy Ion Collider (RHIC) \cite{Adcox:2004mh,Adams:2005dq,Back:2004je,Arsene:2004fa} and CERN Large Hadron Collider (LHC) \cite{Aamodt:2010pa,ATLAS:2011ah,Chatrchyan:2012wg} have played pivotal roles in experimental exploration of the properties of the QGP. The hot and dense medium is quantified as a fluid with extremely small kinetic viscosity by the success of relativistic hydrodynamic description of hadronic momentum distributions and their azimuthal momentum anisotropies \cite{Kolb:2000fha,Schenke:2010rr}. Observed large elliptic flow $v_2$, which can be defined as the second order Fourier coefficient of a transverse momentum ($p_T$) spectrum, is considered as a signature of fluidity because it strongly reflects the spatial anisotropy of the collision system \cite{Poskanzer:1998yz,Ollitrault:1992bk}. 

Comparisons of hydrodynamic simulations and experimental data of $p_T$ spectra \cite{STAR:2006axp,Trainor:2008jp} indicate that low momentum particles below 2-4 GeV are thermalized and participate in the soft medium. Medium to high momentum regions are, with several exceptions such as the hydrodynamic models \cite{Osada:2008sw,Takacs:2019ikb,Kyan:2022eqp} based on nonextensive statistics \cite{Tsallis:1987eu,Tsallis:1999nq}, explained by perturbative QCD approaches and are not considered to constitute the medium. The particle spectra are quantitatively well-described by the combined model in a wide transverse momentum range \cite{Eskola:2002kv,Eskola:2004cr,Eskola:2005ue}. However, most hydrodynamic models assume thermalization of constituent particles at all momenta in the construction of the equation of state and in the calculation of particle production, which can affect the estimation of the observables in collider experiments.

Photons are also abundantly produced in relativistic heavy-ion collisions \cite{Ferbel:1984ef,Alam:1996fd,Cassing:1999es,Peitzmann:2001mz,Gale:2003iz,Rapp:2004qs,Stankus:2005eq,Kapusta:2006pm,David:2006sr,Gale:2009gc,Sakaguchi:2014ewa,David:2019wpt,Monnai:2022hfs}. Inclusive photons are divided into direct photons and decay photons. The former are emitted from the interacting QCD system while the latter are produced after the system is decoupled. Since the QCD medium is transparent in terms of electromagnetic interactions, direct photons are understood as an observable informative of the space-time evolution of the system. Direct photon spectra and flow harmonics, on the other hand, are not completely described by hydrodynamic models \cite{PHENIX:2011oxq,PHENIX:2015igl,ALICE:2018dti}. The situation is known as the \textit{photon puzzle}, and much efforts have been made to solve the issue and gain a comprehensive understanding of photon production in nuclear collisions \cite{Gelis:2004ep,Chatterjee:2005de, Chatterjee:2008tp, Holopainen:2011pd, vanHees:2011vb, Tuchin:2012mf,Basar:2012bp,Bzdak:2012fr,Liu:2012ax,Chatterjee:2013naa,Shen:2013cca,Shen:2013vja,Muller:2013ila,Yin:2013kya, Monnai:2014kqa,Monnai:2014xya,Monnai:2014taa,vanHees:2014ida,Gale:2014dfa, Monnai:2015qha,Linnyk:2015tha,McLerran:2015mda,Paquet:2015lta,Monnai:2015bca,Linnyk:2015rco, Vujanovic:2016anq,Vovchenko:2016ijt,Koide:2016kpe,Dasgupta:2016qkq,Iatrakis:2016ugz,Kim:2016ylr, Fujii:2017nbv,Chatterjee:2017akg,Ayala:2017vex,Dasgupta:2017fns,Dasgupta:2018pjm,Monnai:2018eoh,Dasgupta:2019whr, Ayala:2019jey,Monnai:2019vup,Garcia-Montero:2019kjk,Garcia-Montero:2019vju,Kasmaei:2019ofu, Wang:2020dsr,Dasgupta:2020orj,Churchill:2020uvk,Gale:2021emg,Chatterjee:2021bhz,Schafer:2021slz,Kumar:2021gpf,Chatterjee:2021gwa,Benic:2022ixp,Dasgupta:2022zpa,Fujii:2022hxa,Yazdi:2022cuk,Fujii:2022onx,Ayala:2022zhu,Song:2022ywu,Jia:2022awu,Sethy:2022qic,Sun:2023pil}. The discrepancy between the theoretical estimations and experimental data of direct photon $p_T$ spectra is visible at all momenta while that of direct photon $v_2$ and $v_3$ -- the third order Fourier coefficient known as triangular flow -- is more apparent at higher momenta. One of the difficulties of the puzzle is that the medium-high momentum photons are mainly produced at early times in the time evolution but mechanisms that increase early photon emission reduce $v_2$ and $v_3$ because the flow anisotropy is underdeveloped at the beginning. This implies that a momentum-dependent explanation is required. 

In this study, I develop a hydrodynamic model assuming only low momentum components are thermalized, which may be referred to, in analogy with low frequency light in the visible spectrum, as a red hydrodynamic model for cutting off the contributions of the particles with high momenta. The modified equation of state is estimated based on hadron resonance gas and parton gas models. I show in hydrodynamic model simulations that the medium temperature can be higher at early times but the radial flow development is mostly insensitive to the cutoff. Direct photons are affected by the modification of the medium evolution and the thermal photon emission rate. Numerical simulations indicate that while the effect on direct photon $p_T$ spectra is minimal, differential elliptic flow $v_2(p_T)$ can be sensitive to the momentum dependence of thermalization. 

The paper is organized as follows. In Sec.~\ref{sec:2}, I formulate a relativistic hydrodynamic model with momentum cutoffs for the particles participating in the medium to describe momentum-dependent thermalization. The equation of state is constructed accordingly so that thermodynamic consistency is preserved for macroscopic variables. The results of numerical estimations are shown in Sec.~\ref{sec:3}. Direct photon spectra and elliptic flow are discussed. Conclusions and outlook are presented in Sec.~\ref{sec:4}. I use the natural unit $c = \hbar = k_B = 1$ and the mostly-minus Minkowski metric $g^{\mu \nu} = \mathrm{diag}(+,-,-,-)$ in this paper.

\section{Hydrodynamic model}
\label{sec:2}

The relativistic hydrodynamic model with low momentum components is developed in this section. I consider the situation where the thermal sector is mostly independent of the non-thermal sector, as commonly assumed in a standard hydrodynamic model as the observed particle spectra show a clear separation of description from hydrodynamic to perturbative pictures depending on the momentum scale. Energy-momentum conservation $\partial_\mu T^{\mu \nu} = 0$ provides the equation of motion in the limit of vanishing densities. The modifications of the QCD equation of state and the photon production are discussed in the following sections.

\subsection{Equation of state}

The equation of state is affected by the momentum cutoff for constituent particles. I construct the equation of state based on the hadron resonance gas and parton gas models. 
The grand-canonical partition function $Z_i$ would be expressed as 
\begin{eqnarray}
\ln Z_i = \pm V \int_0^{p_c} \frac{g_i d^3p}{(2\pi)^3} \ln{\bigg[1\pm \exp{\bigg(- \frac{E_i}{T}\bigg) } \bigg]}, 
\end{eqnarray}
in relativistic kinetic theory, which leads to the hydrostatic pressure
\begin{eqnarray}
P &=& \frac{1}{V} \sum_i T \ln Z_i \nonumber \\
&=& \pm T \sum_i \int_0^{p_c} \frac{g_i d^3p}{(2\pi)^3} \ln{\bigg[1\pm \exp \bigg(-\frac{E_i}{T} \bigg) \bigg]}\nonumber \\
&=& \frac{1}{3} \sum_i \int_0^{p_c} \frac{g_i d^3p}{(2\pi)^3} \frac{\mathbf{p}^2}{E_i} \frac{1}{\exp (E_i/T) \pm 1} + \Phi_c, \label{eq:P} 
\end{eqnarray}
where the thermodynamic correction is
\begin{eqnarray}
\Phi_c = \frac{Tp_c^3}{6\pi^2} \sum_i \ln{\bigg[1\pm \exp \bigg(-\frac{E^c_i}{T} \bigg) \bigg]}. \label{eq:Phi} 
\end{eqnarray}
$p_c$ is the cutoff momentum, $V$ is the volume, $g_i$ is the degeneracy factor, $E_i$ is the energy, and $T$ is the temperature. $i$ denotes particle species; $u$, $d$, $s$ quarks and gluons are considered in the QGP phase, and hadron resonances with $u$, $d$, $s$ components with the mass below 2 GeV in the hadronic phase \cite{Tanabashi:2018oca}. 

The hadronic pressure ($P_\mathrm{had}$) and the QGP pressure ($P_\mathrm{QGP}$) are connected as follows \cite{Kyan:2022eqp}:
\begin{equation}
P(T) =
\cases{P_\mathrm{had}(T)& $(T < T_c)$\\
P_\mathrm{QGP}(T)\{1-\exp[-c(T-T_c)]\} & \\
+P_\mathrm{had}(T_c) \exp[-c(T-T_c)] & $(T \geq T_c)$\\}
\end{equation}
where the constant factor is
\begin{eqnarray}
c=\frac{P_\mathrm{had}'(T_c)}{P_\mathrm{QGP}(T_c)-P_\mathrm{had}(T_c)},
\end{eqnarray}
which makes $P(T)$ continuous and smooth at the connecting temperature $T_c$. The pressure of the hadron resonance gas is used up to $T_c$ and then is exponentially damped to the pressure of the parton gas above $T_c$. This prescription has advantages over the conventional hyperbolic connection when the hadronic and QGP pressures have a gap near the crossover region. It puts more emphasis on the hadron resonance gas model, which is known to be consistent with lattice QCD simulations below the crossover temperature, than on the parton gas model. Also, the Cooper-Frye prescription of kinetic freeze-out \cite{Cooper:1974mv} can be used below $T_c$ without suffering from the loss of entropy density when applied to the hydrodynamic model. The difference between the resulting equation of state in the limit of $p_c \to \infty$ and the lattice QCD data \cite{Bazavov:2014pvz,Borsanyi:2013bia} is within 10\% for the temperature range of $0.13 \leq T \leq 0.4$ TeV when $T_c = 0.14$ GeV. Other state variables such as the energy density $e$ and entropy density $s$ are estimated using thermodynamic relations $s = dP/dT$ and $e+P = Ts$.

\begin{figure}[tb]
\begin{center}
\includegraphics[width=3.1in]{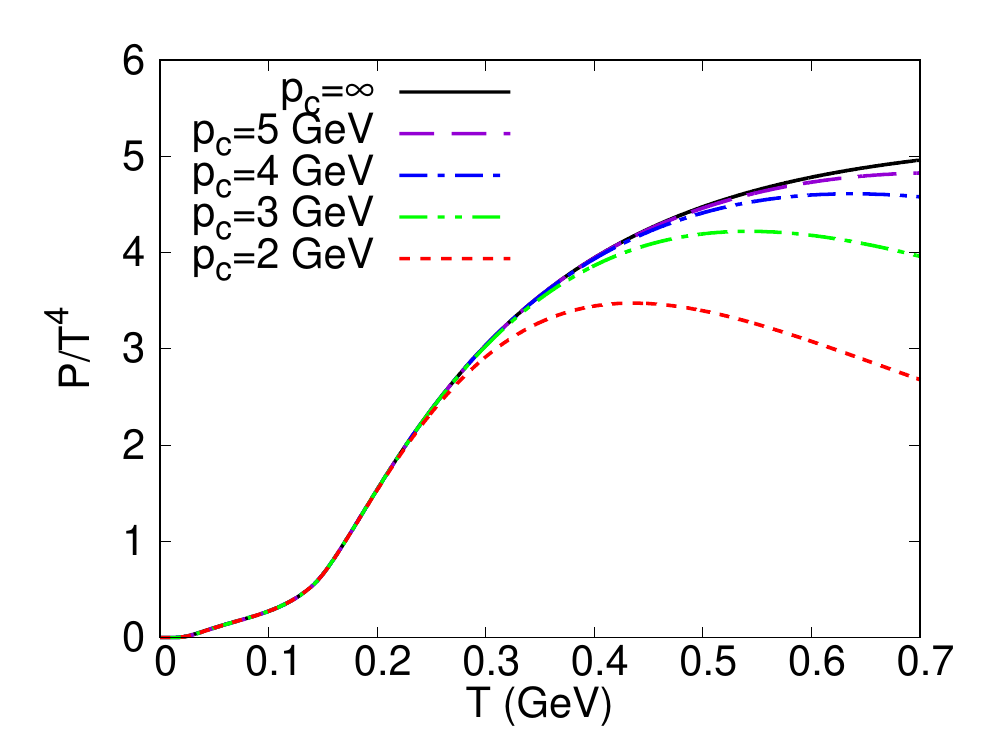}
\caption{(Color online) The equation of state with momentum cutoffs $p_c =$ 5, 4, 3, and 2 GeV (dashed, dash-dotted, dash-double-dotted, and dotted lines, respectively) compared to that without a cutoff (solid line).}
\label{fig:eos}
\end{center}
\end{figure}

The dimensionless pressure $P/T^4$ is shown in Fig.~\ref{fig:eos} for several different values of $p_c$ with $T_c = 0.14$ GeV. One can see that the effects of momentum cutoff are more apparent at higher temperatures and mostly negligible in the hadronic phase for $p_c =$ 2-5 GeV. This implies that the space-time evolution of the medium is affected mostly at early times in nuclear collisions where the temperature is high and the effects on the energy-momentum matching of Cooper-Frye prescription at kinetic freeze-out is small. The effect of cutoff becomes apparent not immediately above $T_c$ because the number of mid-high momentum particles above $p_c$ is still small below around 0.3 GeV for the current parameter range of $p_c$. Thermodynamic quantities are the integration of thermal distribution with weights, and the tail of the distribution has to be large enough for its cutoff to become effective. It should be noted that the pressure $P$ without dimensional normalization monotonously increases as a function of the temperature in all cases, \textit{i.e.}, the entropy density $s(T)$ is positive.

\subsection{Photon production}

Electromagnetic probes are expected to be sensitive to the state of the hot QCD system in early stages of nuclear collisions. In this study, direct photons are considered as the target observable. I consider thermal photons and prompt photons as the conventional sources of direct photons. Additionally, non-thermal contributions from medium-high momentum components are considered because while they are not part of the medium, they could still emit photons. Pre-equilibrium photons and hadron gas photons, which are emitted before and after the medium formation are neglected to focus on the modification of the hydrodynamic stage. The thermal photon contributions outside the freeze-out hypersurface are included \cite{Paquet:2015lta} to partially compensate for the lack of post-equilibrium emissions.

I assume that the momentum-truncated thermal photon emission rate in the QGP phase is expressed based on the small angle prescription \cite{Berges:2017eom, Blaizot:2014jna}, which takes account of the quark pair annihilation and quark-gluon Compton scattering processes, as
\begin{eqnarray}
E \frac{dR^\gamma_\mathrm{QGP}}{d^3p} &=& \sum_f e_f^2 \frac{4}{\pi^2} \alpha_\mathrm{EM} \alpha_s \log \bigg(1+\frac{2.919}{g^2}\bigg) \nonumber \\
&\times& h(p) f_q(p) \int_0^{p_c} \frac{d^3p'}{(2\pi)^3} \frac{1}{p'} [f_g(p') + f_q(p')], \label{eq:QGP} \nonumber \\
\end{eqnarray}
where 
\begin{eqnarray}
h(p)=\frac{1}{2} \bigg[ 1 - \tanh\bigg( \frac{p-p_c}{\Delta p_c} \bigg) \bigg],
\end{eqnarray}
is the hyperbolic factor introduced for a smooth momentum cutoff. $\Delta p_c = 0.2 p_c$ is used. The subscripts $q$ and $g$ denote quarks and gluons. $e_f$ is the charge for the flavor $f$. The couplings are set to $\alpha_s = 0.2$ and $\alpha_\mathrm{EM} = 1/137$. The logarithmic factor is chosen so that the result is consistent with those in Refs.~\cite{Kapusta:1991qp,Baier:1991em} in the thermal limit.

The thermal photon emission rate in the hadronic phase is based on the massive Yang-Mills theory \cite{Gomm:1984at,Song:1993ae} for the gas of light hadrons $\pi$, $K$, $\rho$, $K^*$, and $a_1$ \cite{Turbide:2003si,Heffernan:2014mla,Holt:2015cda}, whose explicit expressions of parametrization can be found in the above-mentioned papers. They are truncated with the factor $h(p)$ to imitate the lack of high $p_T$ contributions.

The total thermal photon emission rate is calculated by connecting the QGP and the hadronic rates as
\begin{eqnarray}
E \frac{dR^\gamma_\mathrm{th}}{d^3p} &=& \frac{1}{2} \bigg[ 1 - \tanh\bigg( \frac{T-T_\textrm{ph}}{\Delta T_\textrm{ph}} \bigg) \bigg] E \frac{dR^\gamma_\mathrm{had}}{d^3p} \nonumber \\
&+& \frac{1}{2} \bigg[ 1 + \tanh\bigg( \frac{T-T_\textrm{ph}}{\Delta T_\textrm{ph}} \bigg) \bigg] E \frac{dR^\gamma_\mathrm{QGP}}{d^3p}, \label{eq:th}
\end{eqnarray}
where $T_\mathrm{ph} = 0.17$ GeV and $\Delta T_\mathrm{ph} = 0.1 T_\mathrm{ph}$. Thermal photon spectra is estimated by replacing $E$ with $p\cdot u$ for taking account of the Lorentz boost before integrating over the space-time volume.

As mentioned earlier, medium-high $p_T$ components might also emit non-thermal photons, which in conventional frameworks are estimated as thermal photons. Owing to the lack of the complete description of this momentum sector, their contribution would be estimated assuming the difference between the thermal rate at $p_c$ and that in the limit of $p_c \to \infty$ represents the high $p_T$ contribution by introducing the effective temperature deduced from the medium temperature,
\begin{eqnarray}
E \frac{dR^\gamma_\mathrm{high}}{d^3p} &=& E \frac{dR^\gamma_\mathrm{th}}{d^3p}(\infty) - E \frac{dR^\gamma_\mathrm{th}}{d^3p}(p_c), \label{eq:high}
\end{eqnarray}
\textit{i.e.}, high $p_T$ photons are conjectured as extra thermal photons and simply assumed to have zero momentum anisotropy because the momentum sector is weakly-coupled. Such a mock-up approach is sometimes employed in direct photon estimations; hadron gas photons from the post-hydrodynamic stage have been estimated by thermal photons outside the particlization hypersurface \cite{Paquet:2015lta} and pre-equilibrium photons have been treated similarly \cite{Shen:2014lye}. While this is a rather simplified approach, it has been suggested that the direct photon yield from a hadronic transport model is roughly approximated by the hydrodynamic approach \cite{Schafer:2021slz}.

In the context of the current study, this method could pose an upper limit for the yield because strongly-interacting media might shine brighter than weakly-coupled ones as the photon production via the $q\bar{q}$ annihilation and $q(\bar{q})g$ scattering processes involves the $\alpha \alpha_s$ factor. It could also provide a lower limit for the elliptic flow because the thermal photons are diluted by the extra photons with zero anisotropy. The prescription keeps the direct photon spectra mostly unchanged because the lack of thermal photons is compensated by non-thermal photons. This might be improved by introducing a transport model designed to describe the medium-high $p_T$ components in future.

Prompt photons are estimated using the parametrization based on the scaling of $p+p$ collision data \cite{Turbide:2003si} as
\begin{eqnarray}
E \frac{dN^\gamma_\mathrm{pr}}{d^3p} &=& 6745 \frac{\sqrt{s}}{(p_T)^5} \frac{N_\mathrm{coll}}{\sigma_{pp}^\mathrm{in}} \frac{\mathrm{pb}}{\mathrm{GeV}^2}.\label{eq:pr}
\end{eqnarray}
$N_\mathrm{coll}$ is the number of collisions and $\sigma_{pp}^\mathrm{in}$ is the inelastic cross section of nucleon-nucleon collisions. They are also assumed to have no azimuthal momentum anisotropy.

\section{Numerical results}
\label{sec:3}
The (2+1)-dimensional inviscid hydrodynamic model \cite{Monnai:2014kqa} is used to numerically elucidate the effects of momentum limits on thermal components. Event-averaged initial conditions for $\sqrt{s_{NN}} = 2.76$~TeV Pb+Pb collisions are constructed using Monte-Carlo Glauber model \cite{Miller:2007ri} at the impact parameter of 4.6 fm to imitate 0-20 \% centrality class events for demonstrative purposes. Full quantitative analyses including event-by-event fluctuation will be discussed elsewhere. The initial time $\tau_\mathrm{hyd} = 0.4$ fm/$c$ and $\sigma_{pp}^\mathrm{in} = 65$ mb are used. The initial energy density at the center of the medium is $e_{0} = 126.6$ GeV/fm$^3$. The kinetic freeze-out temperature of $T_f=140$~MeV is considered to calculate the hadronic yields \cite{Cooper:1974mv} before estimating the resonance decay effects \cite{Sollfrank:1990qz} for normalization. Thermal photon contributions within the extended hypersurface of $T=110$~MeV are taken into account to imitate the effects of hadron gas photons.

\subsection{Space-time evolution of the medium}

\begin{figure}[tb]
\begin{center}
\includegraphics[width=3.1in]{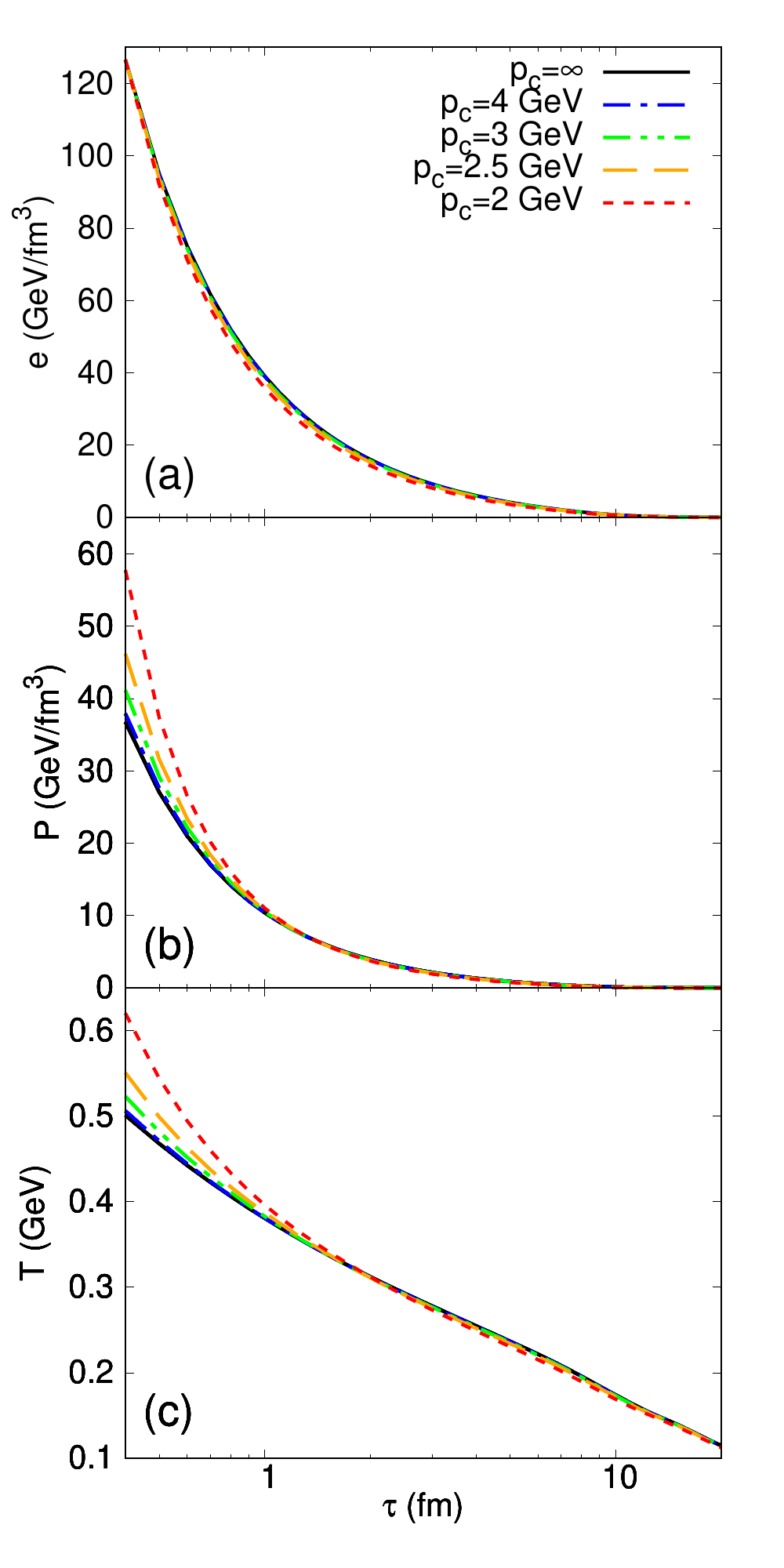}
\caption{(Color online) Time evolution of (a) energy density, (b) pressure, and (c) temperature
at the center of the medium without momentum cutoff (solid line) and with $p_c =$ 4, 3, 2.5, and 2 GeV (dash-dotted, dash-double-dotted, dashed, and dotted lines, respectively).}
\label{fig:evolution}
\end{center}
\end{figure}

The time evolutions of the energy density, pressure, and temperature at the center of the medium without a momentum cutoff and with $p_c =$ 4, 3, 2.5, and 2 GeV are shown in Fig.~\ref{fig:evolution}. The initial energy density is fixed for all cases. One can see that the energy density is mostly insensitive to the momentum cutoff [Fig.~\ref{fig:evolution} (a)]. The pressure, on the other hand, is larger for smaller cutoffs as the deviation from conformality becomes larger [Fig.~\ref{fig:evolution} (b)]. The initial temperature, likewise, is higher for smaller $p_c$ because of the reduction in the effective degrees of freedom for a given temperature [Fig.~\ref{fig:evolution} (c)]. The effects of the momentum cutoff on the pressure and temperature are found to be mostly negligible above $p_c \sim 4$ GeV. The modification becomes small as the medium temperature decreases and the results converge around $\tau \sim 1$ fm/$c$ and $T\sim 0.3$ GeV, which is consistent with the expectations based on the equation of state shown in Fig.~\ref{fig:eos} that the thermodynamic relation between the temperature and the pressure/energy density starts to deviate from that calculated without a momentum cutoff above $T\sim 0.3$ GeV when $p_c \sim 2$ GeV. 
 
\begin{figure}[tb]
\begin{center}
\includegraphics[width=3.1in]{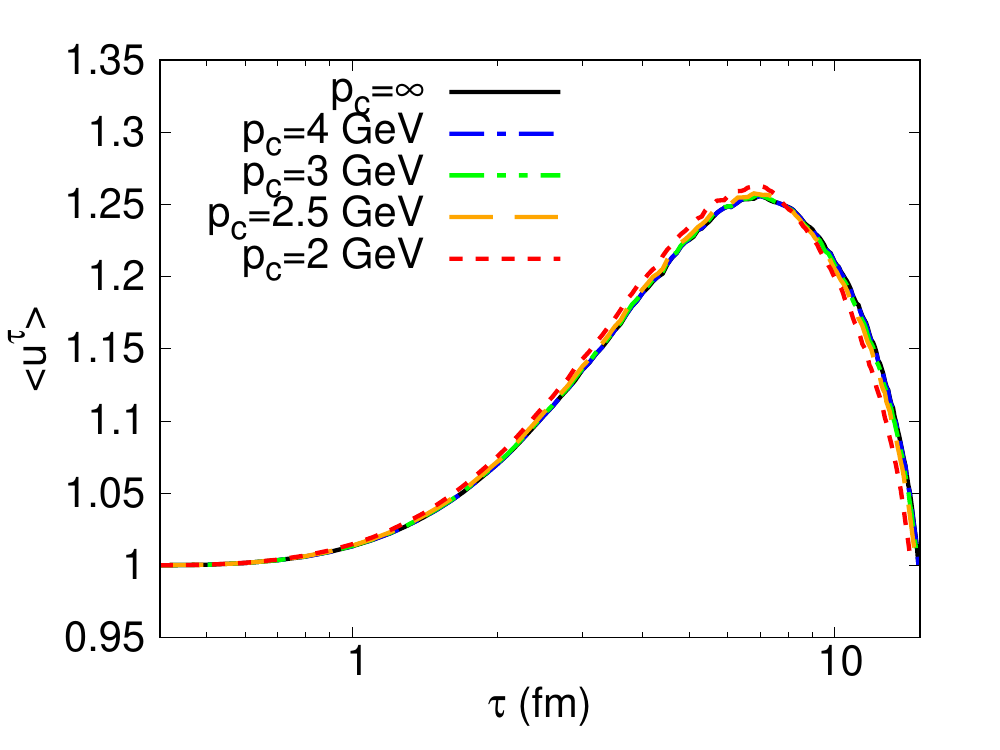}
\caption{(Color online) Time evolution of $\langle u^\tau \rangle$ within the freeze-out hypersurface without momentum cutoff (solid line) and with $p_c =$ 4, 3, 2.5, and 2 GeV (dash-dotted, dash-double-dotted, dashed, and dotted lines, respectively).}
\label{fig:utau}
\end{center}
\end{figure}
 
Figure~\ref{fig:utau} is the time-like flow component $u^\tau$ averaged over the space volume within the freeze-out hypersurface for different $p_c$. The quantity deviates from unity in the presence of the transverse flow because of the normalization condition $u\cdot u = 1$ when the longitudinal flow component $u^{\eta_s}=0$. It increases as a function of time until the peripheral regions start to freeze out. Despite the pressure differences at early times, the flow development is found to be mostly independent of $p_c$, possibly because of the quick convergence of the pressure and its gradients to the values in the $p_c \to \infty$ limit. This implies that it would be difficult to distinguish the $p_c$ scenarios based on the experimental data of hadronic production. 

Radial flow starts to develop after around 1 fm/$c$, suggesting a Bjorken-like longitudinal expansion of the medium \cite{Bjorken:1982qr} at early times. Semi-analytic expressions for the time evolution of the thermodynamic quantities are discussed in \ref{sec:A}.

\subsection{Photon particle spectra}

\begin{figure}[tb]
\begin{center}
\includegraphics[width=3.1in]{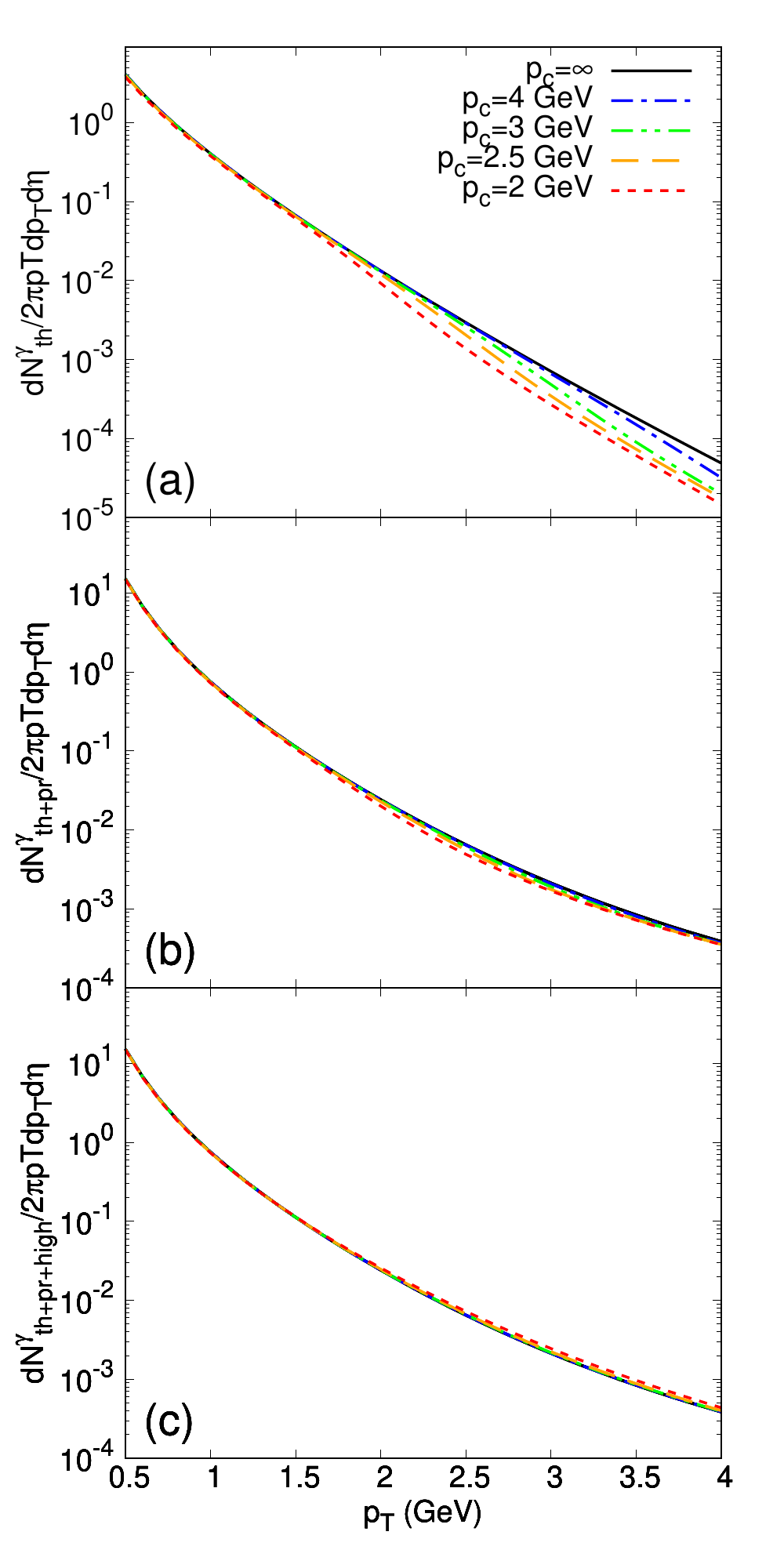}
\caption{(Color online) $p_T$ spectra of (a) thermal photons (b) thermal and prompt photons, and (c) thermal, prompt, and high $p_T$ photons without momentum cutoff (solid line) and with $p_c =$ 4, 3, 2.5, and 2 GeV (dash-dotted, dash-double-dotted, dashed, and dotted lines, respectively).}
\label{fig:pt_spectra}
\end{center}
\end{figure}

Transverse momentum spectra of photons at midrapidity are estimated in numerical simulations. Figure~\ref{fig:pt_spectra} shows $p_T$ spectra of (a) thermal photons, (b) prompt and thermal photons, and (c) direct photons estimated as the sum of prompt, thermal, and high $p_T$ photons. The thermal photon spectrum is visibly reduced above $p_c$ once the cutoff is introduced to the emission rate [Fig.~\ref{fig:pt_spectra} (a)]. The effect competes with that of higher initial temperatures which is supposed to increase the spectrum for smaller $p_c$. When combined with prompt photons, the cutoff effect becomes relatively small and reduces the spectra only around 2-3~GeV because prompt photons become important above around 3~GeV [Fig.~\ref{fig:pt_spectra} (b)]. Figure~\ref{fig:pt_spectra} (c) shows direct photon spectra where non-thermal photons from the high momentum particles are included. It should be noted that the high $p_T$ photons are mimicked by pseudo-thermal photons with medium-high momenta and might be considered as an upper limit of high $p_T$ photon yields as the weakly-coupled sector could emit less photons. While the spectrum is increased by a few percent for smaller $p_c$ because the initial temperature is higher and the thermal photon contribution becomes relatively larger, it is found to be mostly insensitive to the choice of the cutoff momentum.

\subsection{Photon differential elliptic flow}

\begin{figure}[tb]
\begin{center}
\includegraphics[width=3.1in]{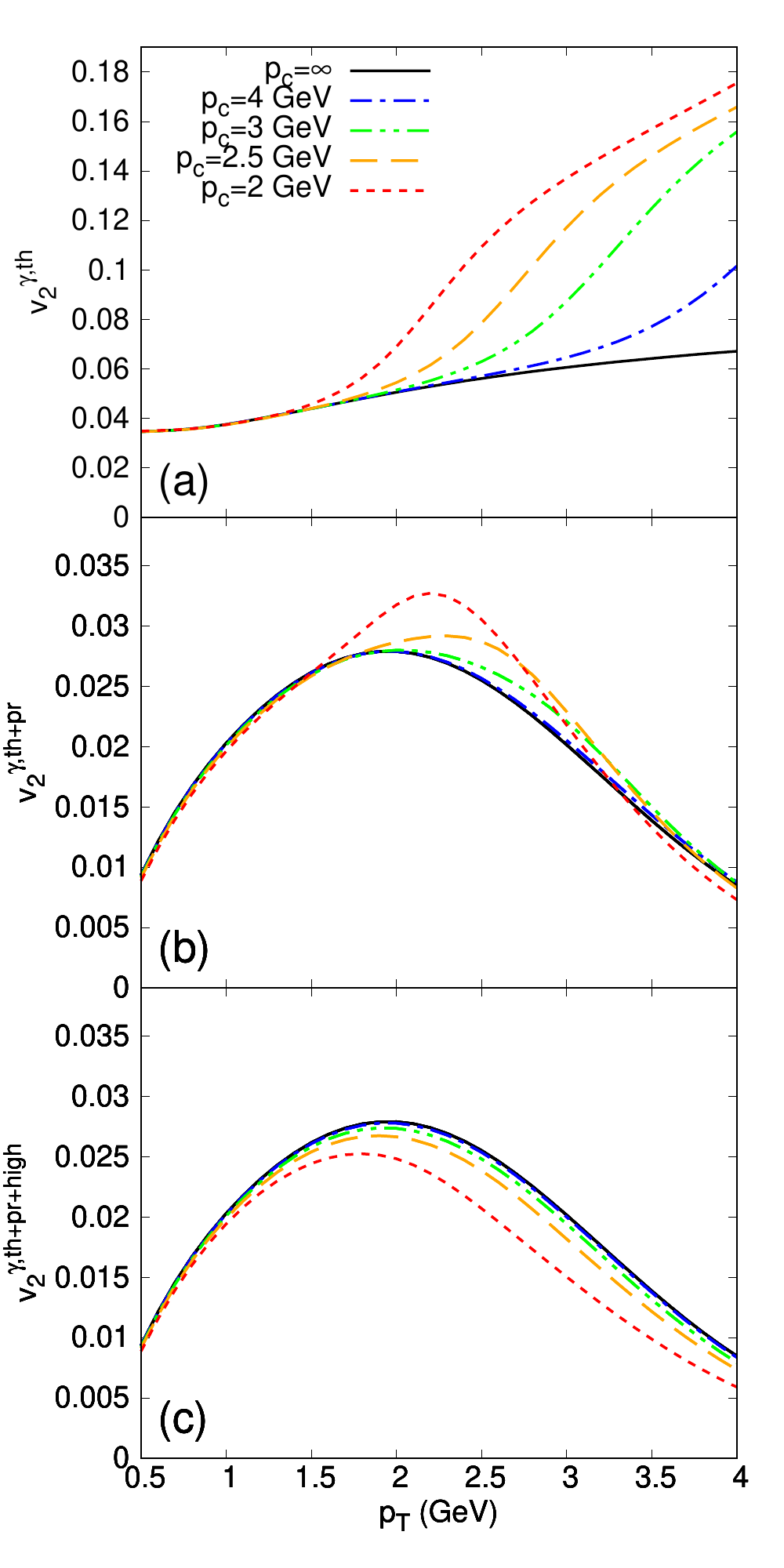}
\caption{(Color online) Differential $v_2$ of (a) thermal photons (b) thermal and prompt photons, and (c) thermal, prompt, and high $p_T$ photons without momentum cutoff (solid line) and with $p_c =$ 4, 3, 2.5, and 2 GeV (dash-dotted, dash-double-dotted, dashed, and dotted lines, respectively).}
\label{fig:v2}
\end{center}
\end{figure}
 
Differential elliptic flow of photons $v^\gamma_2 (p_T)$ at midrapidity are investigated in this section. The azimuthal momentum anisotropy is estimated as 
\begin{eqnarray}
v^\gamma_2 (p_T) &=& \frac{\int d\phi \cos[2(\phi-\Psi)] \frac{dN^\gamma}{d\phi p_T d p_T dy}}{\int d\phi \frac{dN^\gamma}{d\phi p_T d p_T dy}} ,
\end{eqnarray}
where $\phi$ is the azimuthal momentum angle and $\Psi$ is the event plane angle. Thermal photons are the source of anisotropy here because prompt and high $p_T$ photons are assumed to be isotropic.

Figure~\ref{fig:v2} (a) shows thermal photon $v_2$ for different $p_c$. It is visibly enhanced for smaller values of  $p_c$ because the photons above $p_c$ are late time contributions where their momentum is Lorentz-boosted by the radial flow. Note that the number of thermal photons are small above $p_c$. Once prompt photons are added, $v_2$ is still enhanced by the momentum cutoff around $p_T \sim$ 2-3 GeV when $p_c \sim$ 2-3 GeV but is suppressed above those momenta [Fig.~\ref{fig:v2}~(b)]. The contributions of high $p_T$ photons with zero anisotropy further tend to reduce $v_2$ [Fig.~\ref{fig:v2}~(c)]. Compared with the thermal limit $p_c \to \infty$, the upper bound on the momenta of constituent particles might decrease the overall momentum anisotropy because (i) the initial temperature is higher and the number of early thermal photons with underdeveloped anisotropy increases and (ii) the non-thermal contributions above $p_c$ do not have anisotropy. One should keep in mind that the current estimation of high $p_T$ photons is based on a pseudo-thermal photon model and such photons could be produced less in experiments, that direct photon $v_2$ might be less affected by the dilution of non-thermal photons. The results suggest that whether the elliptic flow is enhanced or reduced could depend on non-thermal photons from the medium-high $p_T$ sector. It is also implied that direct photon $v_2$ might be an observable sensitive to the momentum dependence of thermalization.

It should be noted that the results are shown to demonstrate the qualitative behavior of the photon observables caused by the momentum cutoff and cannot be compared quantitatively to experimental data because of the lack of event-by-event fluctuations and viscous modifications, which would affect flow harmonics and, to smaller extent, $p_T$ spectra. Those effects, on the other hand, are expected to be roughly orthogonal to the the cutoff effects and can be studied independently.

\section{Conclusion and outlook}
\label{sec:4}

I have developed the relativistic hydrodynamic model of heavy-ion collisions where the medium is constituted by low momentum components because the experimental data of high-energy heavy-ion collisions suggest that the system is thermalized only up to $p_T \sim$ 2-4 GeV. The QCD equation of state has been constructed using the hadron resonance gas and parton gas models by introducing the upper limit $p_c$ on the momenta of constituents. It has been found that the effect of the momentum limit on the hydrostatic pressure is visible only in the QGP phase for the parameter range of $p_c =$ 2-5 GeV and becomes small for larger $p_c$. 

Numerical hydrodynamic simulations of heavy-ion collisions have indicated that the effects of the momentum cutoff on the thermodynamic variables such as the temperature and pressure are mostly limited to early times before $\tau \sim 1$ fm/$c$ when the initial time is set as $\tau_\mathrm{hyd} =$ 0.4~fm/$c$. The radial flow development is thus not much affected by the choice of $p_c$. It is also consistent with the observation that the effective temperature that characterizes the radial expansion in heavy-ion collisions at RHIC and LHC are 0.2-0.3 GeV \cite{Monnai:2017cbv}, because the equations of state at $p_c =$~2-5~GeV have been shown to converge below 0.3 GeV.
 
The sensitivity of direct photons to the cutoff momentum has been investigated numerically. Particle spectra of thermal photons have been found to decrease above $p_T \sim p_c$. Adding prompt photons make the $p_T$ spectra mostly insensitive to the cutoff because they dominate the spectra above $p_T \sim 3$ GeV where thermal photons are missing. When the conjectured high $p_T$ photons are added, $p_T$ spectra would be slightly enhanced but remain mostly unaffected. Elliptic flow of thermal photons are enhanced above $p_c$ because the small number of high-momentum thermal photons with relatively large azimuthal momentum anisotropy are produced through the Lorentz boost of radial expansion. $v_2$ of thermal and prompt photons combined are still enhanced around $p_T \sim$ 2-3 GeV compared with that of the $p_c \to \infty$ case. If one assumes that non-hydrodynamic high $p_T$ components emit photons with no anisotropy, elliptic flow can be reduced for smaller values of $p_c$ depending on the multiplicity of the high $p_T$ photon emission because those photon may have zero anisotropy. Early thermal photons with small momentum anisotropy would also be increased owing to the higher initial temperature in this case.

The photon puzzle indicates that direct photon $v_2$ might need to be larger than that calculated in conventional methods. This might imply that high $p_T$ photons should not be abundant in the framework of the low-momentum hydrodynamic model. Direct photon $v_2$ might be informative of the momentum dependence of thermalization in heavy-ion collisions as they retain the information about the early stages of hydrodynamic evolution.

Future prospects include the introduction of viscosity and event-by-event fluctuations for qualitative and systematic comparisons with experimental data. It could be interesting to obtain more quantitative analytical solutions of hydrodynamic evolution in such situations \cite{Gubser:2010ui,Hatta:2014upa,Hatta:2014jva,Hatta:2015era}. Also, non-thermal photons from medium-high momentum components might be more quantitatively described by using a weak-coupling picture such as the partonic and hadronic transport models. One may simulate the low momentum sector by  a hydrodynamic model and the high momentum sector by a transport model simultaneously and consider the exchange of energy and momentum between the sectors at each space-time point. It would also be interesting to consider the time dependence of $p_c$ because particles would exchange momentum elastically and inelastically. The model may be extended by introducing finite chemical potentials to elucidate the momentum dependence of thermalization at beam energy scan energies.

\ack
The author is grateful for the valuable discussion with M.~Kitazawa, K.~Murase, A.~Ohnishi, and H.~Suganuma during a seminar at Yukawa Institute for Theoretical Physics, Kyoto University.

\appendix 

\section{Semi-analytical expressions}
\label{sec:A}

One can obtain effective expressions for the time evolution of thermodynamic quantities at early times. The equation of motion can be decomposed into
\begin{eqnarray}
D e &=& -(e+P) \nabla_\mu u^\mu, \\
(e+P) D u^\mu &=& \nabla^\mu P,
\end{eqnarray}
where the time-like and space-like derivatives are $D = u^\mu \partial_\mu$ and $\nabla_\mu = \partial_\mu - u_\mu D$. Figure~\ref{fig:utau} suggests that the flow can be approximated by the Bjorken flow $u^\tau = 1$ before 1~fm/$c$. Then the equation of motion reduces to 
\begin{equation}
\frac{de}{d\tau} = -\frac{e+P}{\tau}.
\end{equation}
If the non-conformal pressure is expressed in terms of the energy density as 
\begin{equation}
P(e)=c_0+c_1e,
\label{eq:peff}
\end{equation}
the analytic solution is 
\begin{equation}
e(\tau) = \frac{[c_0+e_0(1+c_1)](\tau_\mathrm{hyd}/\tau)^{1+c}-c_0}{1+c_1}.
\label{eq:eeff}
\end{equation}
It should be noted that the expression of the type (\ref{eq:peff}) would be valid only in a limited energy density range because the pressure could become negative for a negative offset parameter $c_0$.

The results of numerical fits to the equation of state $P(e)$ between $30 \leq e \leq 126.6$ GeV/fm$^3$, which would be relevant for the time evolution before 1~fm/$c$ in the numerical simulations in Sec.~\ref{sec:3}, are summarized in Table~\ref{table:1}. The expressions~(\ref{eq:peff}) and (\ref{eq:eeff}), supplemented with those parameters, give a reasonable description of the time evolution of the pressure and energy density before 1 fm/$c$.

\begin{table}[htb]
\begin{center}
\begin{tabular}{c|c|c|c|c|c}
\hline \hline
$p_c$ & 2 & 2.5 & 3 & 4 & $\infty$  \\ \hline
$c_0$ (GeV/fm$^3$) & $-7.950$ & $-4.541$ & $-2.755$ & $-1.806$ & $-1.365$ \\ \hline
$c_1$ & 0.506 & 0.393 & 0.342 & 0.311 & 0.300 \\ \hline
\hline
\end{tabular}
\caption{Fitting parameters for the equation of state with finite momentum cutoff.}
\label{table:1}
\end{center}
\end{table}

$c_1$ can be interpreted as the effective sound velocity in the given energy density range. The results suggest that it can increase as the momentum cutoff $p_c$ is lowered.

\section*{References}

\bibliography{low_p}

\end{document}